\documentstyle[12pt]{article}
\newlength{\mathspace}
\def\h #1{\hat{#1}}
\def\b #1{\bar{#1}}
\def\np#1{ Nucl. Phys. B#1}
\def\pr#1    { Phys. Rev. D#1 }
\def\pl#1{ Phys. Lett. B#1}

\def\ijmp#1  { Int. Jour. Mod. Phys. A#1 }
\def\mpl#1   { Mod. Phys. Lett. A#1 }
\def\begineq{\begin{equation}}
\def\endeq{\end{equation}}
\def\eqabegin{\begin{eqnarray}}
\def\eqaend{\end{eqnarray}}
\def\nn{\nonumber}

\begin{document}
\baselineskip=0.7cm
\setlength{\mathspace}{2.5mm}



\begin{titlepage}

    \begin{normalsize}
     \begin{flushright}
                 SINP-TNP/97-08\\
                 hep-th/9706165\\
     \end{flushright}
    \end{normalsize}
    \begin{LARGE}
       \vspace{1cm}
       \begin{center}
         {SL(2, Z) Multiplets of Type II Superstrings in $D < 10$}\\ 
         
       \end{center}
    \end{LARGE}

  \vspace{5mm}

\begin{center}
           
             \vspace{.5cm}

            Shibaji R{\sc oy}
           \footnote{E-mail address: roy@tnp.saha.ernet.in} 

                 \vspace{2mm}

{\it Saha Institute of Nuclear Physics}\\
        {\it 1/AF Bidhannagar, Calcutta 700 064, India}\\

      \vspace{2.5cm}

    \begin{large} ABSTRACT \end{large}
        \par
\end{center}
 \begin{normalsize}
\ \ \ \
It has been shown recently that the toroidally compactified type IIB string
effective action possesses an SL(2, R) invariance. Using this symmetry
we construct an infinite family of macroscopic string-like 
solutions permuted by SL(2, Z) group for type II superstrings in 
$4 \leq D < 10$. These solutions, which formally look very similar 
to the corresponding
solutions in $D = 10$, are characterized by two relatively
prime integers corresponding to the `electric' charges associated with
the two antisymmetric tensor fields of the strings. Stability of these
solutions is discussed briefly in the light of charge conservation and
the tension gap equation.
\end{normalsize}

\end{titlepage}
\vfil\eject

It is well-known that the equations of motion of type IIB supergravity
theory are invariant under an SL(2, R) group known as the supergravity
duality group [1]. A discrete subgroup of this 
group has been conjectured to
be the exact symmetry of the full quantum type IIB string theory [2]. 
As
the string theory coupling constant transforms non-trivially under
this SL(2, R) transformation [3,4,5], this symmetry is 
non-perturbative. So, in
general it is not possible to prove this conjecture in the perturbative
framework of string theory. A strong evidence in favor of this conjecture
has been given by Schwarz [4] when he showed that certain BPS saturated 
macroscopic string-like solutions of type IIB string theory form an SL(2, Z)
multiplet. These solutions, when characterized by two relatively prime 
integers corresponding to the charges associated with the two antisymmetric
gauge fields (from NS-NS and R-R sectors), 
are stable and do not decay further
into strings with lower charges [6]. The tensions as well as the charges
associated with the strings have been shown to be given by SL(2, Z)
covariant expressions.

It has been shown recently at the level of low energy effective action
that this SL(2, R) invariance of the type 
IIB theory survives the toroidal compactification [7,8]. In fact, 
this is not
surprising since a symmetry in a higher dimensional theory should become
a part of the bigger symmetry in the lower dimensional theory, although
in this case, it requires quite non-trivial calculation to prove the 
invariance. So, eventhough SL(2, R) remains a symmetry group of type II
theory  in $D < 10$, the whole symmetry group, known
as the U-duality group [1,2] (E$_9$, E$_8$, E$_7$, E$_6$, 
SO(5, 5), SL(5),
SL(3) $\times$ SL(2), SL(2) $\times$ SO(1, 1) for $D = 2, 3, 4, 5, 6, 7,
8, 9$ respectively) is much bigger and contains both the T-duality group
[9] (O($10 - D$,$10 - D$) in $D$-dimensions) and the 
S-duality group [10,11] (SL(2, R))
as the subgroup. In this paper, we will make use of the SL(2, R) invariance
of type II theory in $D$-dimensions for $4 \leq D <$ 10, to construct
the SL(2, Z) multiplets of macroscopic string-like solutions. Since in 
ten dimensions an SL(2, Z) multiplet of string-like solutions in type IIB
theory has already been shown to exist by Schwarz, 
it is not difficult to understand
that such solutions should also exist in $4 \leq D <$ 10 by direct
dimensional reduction. For $D \leq$ 8 there must exist more string-like
solutions which should form multiplets of bigger symmetry group of type
II theory. Since we are not restricting ourselves to any particular
space-time dimensionality, we construct the multiplets of string-like
solutions belonging to only a subgroup of this bigger symmetry group.
Note that unlike the T-duality group, the SL(2, R) symmetry remains the
symmetry group of type II string theory in any $D<10$. Moreover, since
the T-duality group and the S-duality group act differently on the
background fields, they do not commute with each other and therefore
it is quite non-trivial to see how they combine to form the bigger
symmetry group, namely, the U-duality group. For exapmle, in six
dimensions it has been shown by Sen and Vafa [12] how the complete
U-duality group SO(5,5; Z) arise as a consequence of the T-duality
symmetry O(4,4; Z) and the SL(2, Z) symmetry in type II theory. The
transformation rules for the various background fields have also been
obtained by non-trivial manipulations. But it should be emphasized that
there is no general rule to find the U-duality symmetry and therefore
we consider the non-perturbative subgroup of this bigger symmetry group.
As in the ten dimensional case the string-like solutions constructed
in this paper form SL(2, Z) multiplets as their charges
as well as the tensions are found to be given by SL(2, Z) covariant
expressions. Since these string states are BPS saturated, this gives a 
strong evidence that the compactified theories also possess an exact
SL(2, Z) invariance.
The solutions constructed here will be characterized, as their counterpart
in ten dimensions, by
two relatively prime integers and will be shown to form a stable spectrum
as they are prevented from decaying into other states by the charge 
conservation and the tension gap relation.

The $D$-dimensional effective action common to all string theories has the 
from:
\begineq
\tilde {S}_D = \frac{1}{2 \kappa^2} \int\,d^D x \sqrt {-G} e^{-2 \phi}
\left(R + 4 \partial_\mu \phi \partial^\mu \phi - \frac{1}{12} H_{\mu\nu
\lambda}^{(1)} H^{(1)\,\mu\nu\lambda}\right)
\endeq
where $G = ({\rm det}\, G_{\mu\nu})$, $G_{\mu\nu}$ being the string metric,
$R$ is the scalar curvature associated with $G_{\mu\nu}$, $\phi$ is the
dilaton and $H_{\mu\nu\lambda}^{(1)}$ is the field strength associated with
the Kalb-Ramond antisymmetric tensor field $B_{\mu\nu}^{(1)}$. These are
the massless modes which couple
to any string theory. By a conformal scaling of the metric
\begineq
G_{\mu\nu} = e^{\frac{4}{D-2} \phi} g_{\mu\nu}
\endeq	
we can rewrite the action in the Einstein frame as follows:
\begineq
{S}_D = \frac{1}{2 \kappa^2} \int\,d^D x \sqrt {-g} 
\left[R - \frac{4}{D-2} \partial_\mu \phi \partial^\mu \phi - \frac{1}{12} 
e^{-\frac{8}{D-2} \phi} H_{\mu\nu
\lambda}^{(1)} H^{(1)\,\mu\nu\lambda}\right]
\endeq
Here $R$ now represents the scalar curvature associated with the 
canonical metric
$g_{\mu\nu}$. The above $D$-dimensional action is precisely the one 
studied by Dabholkar et. al. [13] to construct the 
macroscopic string-like
solutions. (We will show later that the $D$-dimensional type II action
reduces to the action (3) in a particular limit.)
Note that we have not rescaled the dilaton as done in ref.[13].
The solution is given by the following metric and the other field
configurations,
\begineq
ds^2 = A^{-\frac{D-4}{D-2}}\left(-dt^2 + (dx^1)^2\right) + A^{\frac{2}{D-2}}
\delta_{ij} dx^i dx^j
\endeq
Here $i,\,j = 2, \ldots, D-1$. $A$ is a function (whose explicit form
is given below) of radial coordinate $r$ only, where, 
$r^2 \equiv \delta_{ij} x^i x^j$. 
It is clear from (4) that the string is aligned to 
$x^1$. The only non-zero component of the
antisymmetric tensor field is
\begineq
B_{01}^{(1)} = A^{-1} = e^{2\phi}
\endeq
It should be noted that the expression (5) is formally 
independent of the
dimensionality of the theory and this is the reason for the formal
similarity of the string-like solutions in $D = 10$ and $D < 10$ as
we will see later. Of course, the information about the 
dimensionality resides in the function $A$. The function $A(r)$ 
is given by,
\begineq
A(r) = \left\{\begin{array}{ll} 1 + \frac{Q}{(D-4) r^{D-4} \Omega_{D-3}} 
& \qquad{\rm for}\quad D > 4\\
1 - \frac{Q}{2\pi} \log r & \qquad{\rm for}\quad D = 4
\end{array}\right.
\endeq
Here $Q$ is the `electric' charge associated with the antisymmetric tensor
field $B_{\mu\nu}^{(1)}$ carried by the string and is defined as,
\begineq
Q = \int_{S^{D-3}} \ast e^{-\frac{8}{D-2} \phi} H^{(1)}
\endeq
Here $\ast$ denotes the Hodge dual in terms of the canonical metric and
the integral is to be evaluated over ($D - 3)$-dimensional unit sphere
$S^{D-3}$ surrounding the string. Also $\Omega_{D-3}$ represents the volume
of the unit ($D - 3)$-dimensional sphere given by $\Omega_{D-3} = 
\frac{ 2 (\pi)^{\frac{D-2}{2}}}{\Gamma[\frac{1}{2} (D-2)]}$.

Recall that in ref.[13], the field equations and their solution (6) were
obtained by coupling the supergravity action $S_D$ to a macroscopic string
source $S_\sigma$ given by
\begineq
S_\sigma = -\frac{T}{2} \int\,d^2\sigma \left(\sqrt{\gamma} 
\gamma^{\alpha\beta} \partial_\alpha X^\mu \partial_\beta X^\nu g_{\mu\nu}
e^{\frac{4}{D-2} \phi} + \epsilon^{\alpha\beta} \partial_\alpha X^\mu
\partial_\beta X^\nu B_{\mu\nu}^{(1)}\right)
\endeq
where $T$ is the string tension and $\gamma_{\alpha\beta}$ is the world-sheet
metric. So, the $D$-dimensional string-like solution in (4) is not `solitonic'
in the strict sense as the metric shows a curvature singularity at $r = 0$
and also the field equation $\nabla^2 A$ has a delta function singularity
at that point [11]. By evaluating the `electric' 
charge explicitly using Eq.(7)
we can derive the relation between the charge and string tension as $Q =
2 \kappa^2 T$. We would like to point out that the supergravity action (3)
has a manifestly SL(2, R) invariance by which the fields in (3) can be
rotated to convert the action to the $D$-dimensional type II action including
the R-R terms. Since we have already obtained the toroidal compactification
of type IIB theory in ref.[8], we will show how starting 
from $D$-dimensional
type II action we can obtain the action (3) of Dabholkar et. al.

It is well-known that the equations of motion of type IIB supergravity
theory can not be obtained from a covariant action [14] because of the 
presence
of a four-form gauge field with the self-dual field strength in the spectrum.
This gauge field couples to a self-dual three-brane which can give rise
to string solution in $D\leq$ 8. But, we are not going to consider this
type of string solution and set the corresponding field-strength $F_5$ to 
zero. There are also magnetically charged string solution for type II theory
in $D\leq  6$, but since we are not restricting ourselves 
to any particular dimensionality
we will not consider those kinds of solutions also. Now as we set $F_5 = 0$,
the type IIB equations of motion can be derived from the following covariant
action\footnote[1]{Here $\mu,\nu,\ldots = 0, 1, \ldots,D-1$ are the 
space-time indices and
$m,n,\ldots = D,\ldots,9$ are the internal indices. 
The ten dimensional objects are
denoted with a `hat'. Note that we are using slightly different notation than
in ref.[8].}:
\eqabegin
\tilde{S}_{10}^{\rm IIB} &=& \frac{1}{2\kappa^2}
\int\,d^{10} \h x \sqrt {- \h {G}} \left[ e^{-2 \h {\phi}}
\left(\h {R} + 4 \partial_{\h \mu} \h {\phi} \partial^{\h \mu} 
\h {\phi}
- \frac{1}{12} \h {H}_{\h \mu \h \nu \h \lambda}^{(1)} \h {H}^{(1)\, \h \mu
\h \nu \h \lambda}\right)\right.\nn\\
& &\qquad \left. - \frac{1}{2} \partial_{\h \mu} \h {\chi} \partial^
{\h \mu}
\h {\chi} -\frac{1}{12}\left(\h {H}_{\h \mu \h \nu \h \lambda}^{(2)} 
+ \h {\chi} \h {H}_{\h \mu \h \nu \h \lambda}^{(1)}\right)\left(
\h {H}^{(2)\, \h \mu \h \nu \h \lambda} + \h {\chi} \h {H}^{(1)\,
\h \mu \h \nu \h \lambda}\right)\right]
\eqaend
Here the metric $\h {G}_{\h \mu \h \nu}$, the dilaton $\h \phi$ and the
antisymmetric tensor $\h{B}_{\h \mu \h \nu}^{(1)}$ (with $\h {H}^{(1)}
= d \h {B}^{(1)}$) represent the massless modes in the NS-NS sector of 
type IIB theory. Also the scalar $\h \chi$ and $\h {B}_{\h \mu \h \nu}^{(2)}$
(with $\h {H}^{(2)} = d \h {B}^{(2)}$) represent the massless modes in
the R-R sector. We have already studied the dimensional reduction of this
action in ref.[8]. The reduced action takes the form:
\eqabegin
& &\frac{1}{2\kappa^2} \int\, d^D x \sqrt{-{G}}\left[e^{-2 \phi} 
\left(
R + 4 \partial_\mu \phi \partial^\mu \phi - \frac{1}{4}
G_{mn} F_{\mu\nu}^{(3)\,m} F^{(3)\,\mu\nu,\,n} + \frac{1}{4}
\partial_\mu G_{mn}\partial^\mu G^{mn}\right.\right.\nn\\
& &\qquad\qquad\left. 
-\frac{1}{4}
G^{mp} G^{nq}\partial_\mu B_{mn}^{(1)} \partial^\mu B_{pq}^{(1)}
  -\frac{1}{4}
G^{mp} H_{\mu\nu\,m}^{(1)} H^{(1)\,\mu\nu}_{\,\,\,\,p} - \frac{1}{12}
H_{\mu\nu\lambda}^{(1)} H^{(1)\,\mu\nu\lambda}\right)\nn\\
& &\qquad\qquad -\frac{1}{2} \Delta \partial_\mu \chi \partial^\mu
\chi
 -\frac{1}{4} \Delta G^{mp} G^{nq}\left(\partial_\mu 
B_{mn}^{(2)} + \chi \partial_\mu B_{mn}^{(1)}\right)\left(\partial^\mu
B_{pq}^{(2)} + \chi \partial^\mu B_{pq}^{(1)}\right)\nn\\
& &\qquad\qquad\qquad -\frac{1}{4}\Delta G^{mp}\left(H_{\mu\nu\,m}
^{(2)} + \chi
H^{(1)}_{\mu\nu\,m}\right)\left(H^{(2)\,\mu\nu}_{\,\,\,\,p} + \chi
H^{(1)\,\mu\nu}_{\,\,\,\,p}\right)\nn\\
& &\qquad\qquad\qquad\left. -\frac{1}{12}\Delta\left(H^{(2)}_{\mu\nu\lambda} +
\chi H^{(1)}_{\mu\nu\lambda}\right)\left(H^{(2)\,\mu\nu\lambda} +
\chi H^{(1)\,\mu\nu\lambda}\right)\right]
\eqaend
where the definitions and the reduced forms of various gauge fields are:
\eqabegin
\h G_{\h \mu \h \nu} &\longrightarrow& \left\{\begin{array}{l}
\h G_{mn}\,\,=\,\,G_{mn}\\
G_{\mu m}\,\,=\,\,\h G_{\mu m}\,\,=\,\, A_\mu^{(3)\,n} G_{mn}\\
\h G_{\mu\nu}\,\,=\,\, G_{\mu\nu} + G_{mn} A_\mu^{(3)\,m} A_{\nu}
^{(3)\,n}\end{array}\right.\\
\h {\phi}\,\,&=&\,\, \phi + \frac{1}{2} \log \Delta, \quad {\rm where}
\qquad \Delta^2 = ({\rm det}\,G_{mn})\\
\h \chi\,\,&=&\,\, \chi\\
\h {B}_{\h \mu \h \nu}^{(i)} &\longrightarrow& \left\{\begin{array}
{l} B_{\mu m}^{(i)}\,\,=\,\, A_{\mu m}^{(i)}\,\,=\,\, \h {B}_{\mu m}^{(i)}
+ B_{mn}^{(i)} A_{\mu}^{(3)\,n}\\
B_{\mu\nu}^{(i)}\,\,=\,\, \h {B}_{\mu\nu}^{(i)} + A_\mu^{(3)\,m} 
A_{\nu m}^{(i)} - A_\nu^{(3) m} A_{\mu m}^{(i)} - A_\mu^{(3) m} B_{mn}^{(i)}
A_\nu^{(3) n}\end{array}\right.
\eqaend
where $i = 1, 2$. The corresponding field-strengths are given below:
\eqabegin
H_{\mu mn}^{(i)} &=& \h {H}_{\mu mn}^{(i)}\,\,=\,\,\partial_\mu B_{mn}^{(i)}\\
H_{\mu\nu m}^{(i)} &=& F_{\mu\nu m}^{(i)} - B_{mn}^{(i)} F_{\mu\nu}^{(3)\,n}
\eqaend
where $F_{\mu\nu\,m}^{(i)} = \partial_\mu A_{\nu\,m}^{(i)} - \partial_\nu
A_{\mu\,m}^{(i)}$ and $F_{\mu\nu}^{(3)\,m} = \partial_\mu A_\nu^{(3)\,m}
- \partial_\nu A_\mu^{(3)\,m}$ and finally, 
\begineq
H_{\mu\nu\lambda}^{(i)}\,\,=\,\,\partial_\mu B_{\nu\lambda}^{(i)} - 
F_{\mu\nu
}^{(3)\,m} A_{\lambda\,m}^{(i)} + {\rm cyc.\,\,\, in\,\,\, \mu\nu\lambda
}\endeq
The reduced action (10) was shown in ref.[8] to have an SL(2, R) invariance
which can be better understood by rewriting the action in the Einstein frame.
The metric in the Einstein frame is related with the string metric as
given in Eq.(2). Using this, the action (10) in the Einstein frame takes the
following form:
\eqabegin
& &\frac{1}{2\kappa^2} \int\, d^D x \sqrt{-g}\left[ 
R -\frac{1}{2} \partial_\mu \tilde{\phi} \partial^\mu 
\tilde{\phi} -\frac{1}{2} e^{2 \tilde{\phi}} \partial_\mu \chi
\partial^\mu \chi + \frac{1}{8} \partial_\mu \log {\b {\Delta}}
\partial^\mu \log{\b{\Delta}}\right.\nn\\ 
& &\qquad +\frac{1}{4} \partial_\mu {g}_{mn} \partial^\mu {g}^{mn}
- \frac{1}{4}
{g}_{mn} F_{\mu\nu}^{(3)\,m} F^{(3)\,\mu\nu,\,n}
-\frac{1}{4}(\b {\Delta})^{1/2}
{g}^{mp} {g}^{nq} e^{-\tilde{\phi}}\partial_\mu B_{mn}^{(1)} 
\partial^\mu B_{pq}^{(1)}\nn\\
 & &\qquad
 -\frac{1}{4}(\b{\Delta})^{1/2} {g}^{mp} {g}^{nq} e^{\tilde {\phi}}
\left(\partial_\mu 
B_{mn}^{(2)} + \chi \partial_\mu B_{mn}^{(1)}\right)\left(\partial^\mu
B_{pq}^{(2)} + \chi \partial^\mu B_{pq}^{(1)}\right)\\
& &\qquad -\frac{1}{4}(\b{\Delta})^{1/2} {g}^{mp}\left\{e^{-\tilde{\phi}}
H_{\mu\nu m}^{(1)} H^{(1)\,\mu\nu}_{\,\,\,\,p} + e^{\tilde{\phi}}
\left(H_{\mu\nu\,m}^{(2)} + \chi
H^{(1)}_{\mu\nu\,m}\right)\left(H^{(2)\,\mu\nu}_{\,\,\,\,p} + \chi
H^{(1)\,\mu\nu}_{\,\,\,\,p}\right)\right\}\nn\\
& &\qquad\left. -\frac{1}{12}(\b{\Delta})^{1/2}\left
\{e^{-\tilde{\phi}}
H_{\mu\nu\lambda}^{(1)} H^{(1)\,\mu\nu\lambda} + e^{\tilde {\phi}}
\left(H^{(2)}_{\mu\nu\lambda} +
\chi H^{(1)}_{\mu\nu\lambda}\right)\left(H^{(2)\,\mu\nu\lambda} +
\chi H^{(1)\,\mu\nu\lambda}\right)\right\}\right]\nn
\eqaend
where we have defined $\tilde {\phi} = \phi + \frac{1}{2} \log \Delta$.
Also, $G_{mn} = e^{\frac{4}{D-2} \phi} g_{mn}$ and so, $\Delta =
e^{2 \frac{(10 - D)}{(D - 2)} \phi} \b {\Delta}$ with $(\b {\Delta})^2
= ({\rm det}\, g_{mn})$. If we define the following SL(2, R) matrix
\begineq
{\cal M}_D\,\,\equiv\,\, \left(\begin{array}{cc}
\chi^2 + 
e^{- 2\tilde {\phi}}
&  \chi \\
 \chi  &  1\end{array}\right)\,\,e^{\tilde {\phi}}
\endeq
then the action (18) can be expressed in the manifestly SL(2, R) invariant
form as,
\eqabegin
& &\frac{1}{2\kappa^2}\int\,d^D x \sqrt{-{g}}\left[{R} + \frac{1}{4} 
{\rm tr}\,
\partial_\mu {\cal M}_D \partial^\mu {\cal M}_D^{-1} +
\frac{1}{8} \partial_\mu \log
\b {\Delta} \partial^\mu \log \b {\Delta} + \frac{1}{4} \partial_\mu
{g}_{mn} \partial^\mu {g}^{mn}\right.\nn\\
& &\qquad\qquad\qquad -\frac{1}{4} {g}_{mn} F_{\mu\nu}^{(3)\,m} F^{(3)\,
\mu\nu,\,n} - \frac{1}{4} (\b \Delta)^{1/2} {g}^{mp} {g}^{nq}
\partial_\mu {\cal B}_{mn}^T {\cal M}_D \partial^\mu {\cal B}_{pq}\\
& &\qquad\qquad\qquad\left. -\frac{1}{4} 
(\b \Delta)^{1/2} {g}^{mp} {\cal H}
^T_{\mu\nu\,m} {\cal M}_D {\cal H}^{\mu\nu}_{\,\,\,\,p} - \frac{1}{12}
(\b \Delta)^{1/2} {\cal H}_{\mu\nu\lambda}^T {\cal M}_D 
{\cal H}^{\mu\nu\lambda}\right]\nn
\eqaend
Here we have defined 
${\cal B}_{mn}\,\,\equiv \,\,
 \left(\begin{array}{c} B_{mn}^{(1)}  \\
B_{mn}^{(2)}\end{array}\right)$, ${\cal H}_{\mu\nu\,m}\,\,\equiv \,\,
\left(\begin{array}{c} H_{\mu\nu\,m}^{(1)} \\ H_{\mu\nu\,m}^{(2)} \end{array}
\right)$, and ${\cal H}_{\mu\nu\lambda}\,\,\equiv\,\,\left(\begin{array}{c}
H_{\mu\nu\lambda}^{(1)} \\ H_{\mu\nu\lambda}^{(2)}\end{array}\right)$. 
 The superscript `$T$' denotes the tranpose of a matrix. The action 
(20) is invariant under the following global SL(2, R)
transformation:
\eqabegin
{\cal M}_D &\rightarrow& \Lambda {\cal M}_D \Lambda^T, \qquad {\cal B}_{mn}
\,\,\,\rightarrow\,\,\,(\Lambda^{-1})^T {\cal B}_{mn}\nn\\ 
\left(\begin{array}{c} A_{\mu\,m}^{(1)}\\ A_{\mu\,m}^{(2)}\end{array}\right)
&\equiv& {\cal A}_{\mu\,m}\,\,\,\rightarrow\,\,\,(\Lambda^{-1})^T
{\cal A}_{\mu\,m},\qquad
\left(\begin{array}{c} B_{\mu\nu}^{(1)}\\ B_{\mu\nu}^{(2)}\end{array}\right)
\,\,\equiv\,\, {\cal B}_{\mu\nu}\,\,\,\rightarrow\,\,\,(\Lambda^{-1})^T 
{\cal B}_{\mu\nu}\nn\\
 {g}_{\mu\nu}&\rightarrow& {g}_{\mu\nu},
\qquad {g}_{mn}\,\,\,\rightarrow\,\,\,{g}_{mn},\qquad
{\rm and}\quad A_{\mu}^{(3)\,m}\,\,\,\rightarrow\,\,\,A_\mu^{(3)\,m}
\eqaend
where $\Lambda\,\,=\,\,\left(\begin{array}{cc} a & b\\ c & d\end{array}
\right)$ is the SL(2,R) transformation matrix and $a, b, c, d$ are 
constants satisfying $ad - bc = 1$. Now if we set $G_{mn} = \delta_{mn}$,
$\Delta = 1$, $A_\mu^{(3)\,m} = A_{\mu\,n}^{(i)} = B_{mn}^{(i)} = 0$,
then the action (20) reduces to:
\eqabegin
& &\frac{1}{2\kappa^2}\int\,d^D x \sqrt{-{g}}\left[{R} + \frac{1}{4} 
{\rm tr}\,
\partial_\mu {\cal M}_D \partial^\mu {\cal M}_D^{-1} +
\frac{1}{8} \partial_\mu \log
\b {\Delta} \partial^\mu \log \b {\Delta}\right.\nn\\ 
& &\qquad\qquad\qquad\left. + \frac{1}{4} \partial_\mu
{g}_{mn} \partial^\mu {g}^{mn}
- \frac{1}{12}
(\b \Delta)^{1/2} {\cal H}_{\mu\nu\lambda}^T {\cal M}_D 
{\cal H}^{\mu\nu\lambda}\right]
\eqaend
This action is SL(2, R) invariant under the transformation (21). Note
that both $g_{mn}$ and $\b {\Delta}$ are SL(2, R) invariant. Also,
${\cal M}_D$ in (21) is as given in (19) with $\tilde {\phi}$ replaced
by $\phi$, the $D$-dimensional dilaton as $\Delta = 1$ in this case. Note
also that although we have set $G_{mn} = \delta_{mn}$ and $\Delta = 1$,
but as they are not SL(2, R) invariant, non-trivial values of $G_{mn}$
and $\Delta$ will be generated through SL(2, R) transformation. It can
be easily checked that the SL(2, R) invariant action (22) gets precisely
converted to the effective action (3) considered by Dabholkar et. al.
by setting the R-R fields to zero. Thus, we note that the action (3)
is a special case of the more general type II action (22) and so the
solution (6) is a particular case of a general solution that we are
going to construct.

Our strategy to construct the string-like solution for type II theory
in $D$-dimensions is to start with the action (3) and the solution 
(4)--(6) and then use the SL(2, R) symmetry to rotate the solution 
corresponding to the full type II theory. Since the starting solution
has only $B_{\mu\nu}^{(1)}$ field and the corresponding charge, the 
final solution will have both the fields $B_{\mu\nu}^{(1)}$ and
$B_{\mu\nu}^{(2)}$ and their charges. In order to describe the complete
string solution we also have to specify the asymptotic values of both
$\phi$ and $\chi$ as $r \rightarrow \infty$. Under the transformation
${\cal M}_D \rightarrow \Lambda {\cal M}_D \Lambda^T$ and ${\cal B}_
{\mu\nu} \rightarrow (\Lambda^{-1})^T {\cal B}_{\mu\nu}$, the complex 
scalar field $\lambda = \chi + i e^{- \phi}$ and the ${\cal B}_{\mu\nu}^
{(i)}$ transform as,
\eqabegin
\lambda &\rightarrow& \frac{a\lambda + b}{c\lambda + d}\\
B_{01}^{(1)} &\rightarrow& d B_{01}^{(1)} - c B_{01}^{(2)}\nn\\
B_{01}^{(2)} &\rightarrow& - b B_{01}^{(1)} + a B_{01}^{(2)}
\eqaend
We first construct the solution corresponding to the simplest choice of
$\lambda_0 = i$ (i.e. for $\chi_0 = \phi_0 = 0$) as in ref.[4]. 
Here subscript `$0$' represents the asymptotic value. We also
replace the `electric' charge $Q$ in (6) by $\alpha_{(q_1, q_2)} = 
\sqrt {q_1^2 + q_2^2} Q$. From the relation between charge and string
tension $Q = 2\kappa^2 T$, it is clear that $T$ also has 
to be replaced by $T_{(q_1, q_2)}
= \sqrt {q_1^2 + q_2^2} T$. Using the value of $B_{01}^{(1)} = A_{(q_1, q_2
)}^{-1}$, where $A_{(q_1, q_2)}$ is as given in (6) with $Q$ replaced
by $\alpha_{(q_1, q_2)}$, we can easily calculate the `electric' charges
associated with the transformed fields $B_{01}^{(1)}$ and $B_{01}^{(2)}$
to be $a\sqrt{q_1^2 + q_2^2} Q$ and $c\sqrt{q_1^2 + q_2^2} Q$. 
Demanding that
the charges be quantized, the constants $a, b, c, d$ get completely fixed as
follows:
\begineq
\Lambda = \frac{1}{\sqrt {q_1^2 + q_2^2}}\left(\begin{array}
{cc} q_1 & -q_2\\
q_2 & q_1\end{array}\right)
\endeq
where $q_1$ and $q_2$ are integers. With this $\Lambda$, $B_{01}^{(i)}$
and $\lambda$ are given as
\eqabegin
B_{01}^{(i)} &=& \frac{q_i}{\sqrt{q_1^2 + q_2^2}} A_{(q_1, q_2)}^{-1}\\
\lambda &=& \frac{i q_1 A^{1/2}_{(q_1, q_2)} - q_2}{i q_2 A^{1/2}_
{(q_1, q_2)} + q_1}\,\,\,=\,\,\,\frac{q_1 q_2 \left(A_{(q_1, q_2)} - 1
\right) + i \left(q_1^2 + q_2^2\right) A^{1/2}_{(q_1, q_2)}}{q_1^2 +
q_2^2 A_{(q_1, q_2)}}
\eqaend
Note that the solution formally has the same form as the ten dimensional
solution [4]. Asymptotically for $D > 4$ as $r\rightarrow \infty$, 
$A_{(q_1, q_2)}
\rightarrow 1$ and so, $\lambda \rightarrow i$. Note from Eqs.(4) and
(6), that the metric has a nice asymptotic limit for $D > 4$, but for
$D = 4$ the metric has logarithmic divergence. However, it was shown
in ref.[13] that one can still write the metric in a ``weak-field''
type expansion and calculate the total energy density in the ambient
space. It was found that the energy density outside the string core
($r = 0$) vanishes reflecting the fact that effectively the function
$A_{(q_1,q_2)}$ tends to one as we go away from the string core. 
It should be mentioned
that although originally $\chi = 0$, a non-trivial $\chi$ is generated
through SL(2, R) transformation as can be seen from (27). Similarly,
although we started with a trivial internal metric $G_{mn} = \delta_{mn}$
and $\Delta = 1$, a non-trivial $G_{mn}$ and $\Delta$ will be
generated after the SL(2, R) transformation as given below:
\eqabegin
G_{mn} &=& \left(\frac{q_1^2 + q_2^2 A_{(q_1, q_2)}}{q_1^2 + q_2^2}
\right)^{4/{(D-2)}}\,\delta_{mn}\nn\\
\Delta &=& \left(\frac{q_1^2 + q_2^2 A_{(q_1, q_2)}}{q_1^2 + q_2^2}
\right)^{2 (10-D)/(D-2)}
\eqaend
In deriving (28) we have used the transformation rule of $e^{-\phi}$
obtained in (27) and the fact that $g_{mn} = e^{- \frac{4}{D-2} \phi}
G_{mn}$ is invariant under SL(2, R) transformation.

We next generalize the construction for arbitrary vacuum modulus 
$\lambda_0$. So, we start with an arbitrary value of the charge $\alpha_
{(q_1, q_2)} = \Delta^{1/2}_{(q_1, q_2)} Q$ in (6) and choose,
\eqabegin
\Lambda\,\,\,=\,\,\,\Lambda_1\,\Lambda_2 &=& \left(\begin{array}{cc}
e^{-\phi_0/2} & \chi_0 e^{\phi_0/2}\\
0 & e^{\phi_0/2}\end{array}\right)\,\,\left(\begin{array}{cc}
\cos \theta & - \sin \theta\\
\sin \theta & \cos \theta\end{array}\right)\nn\\
&=& \left(\begin{array}{cc} 
e^{-\phi_0} \cos \theta + \chi_0 \sin \theta & - e^{-\phi_0} \sin \theta
+ \chi_0 \cos \theta\\
\sin \theta & \cos \theta\end{array}\right)\,e^{\phi_0/2}
\eqaend
Note that $\Lambda_2$ is the most general SL(2, R) matrix which preserves
the vacuum modulus $\lambda_0 = i$ and $\Lambda_1$ is the SL(2, R) matrix 
which transforms it to an arbitrary value $\lambda = \lambda_0$. We point
out that the charge quantization condition in the previous case fixed the
value of $\cos \theta$ and $\sin \theta$ to have the particular 
form given in
(25). In the present case as the matrix $\Lambda$ is different, charge
quantization condition will yield different values of $\cos \theta$
and $\sin \theta$. Using (29), we find the `electric' charges associated
with $B_{01}^{(1)}$ and $B_{01}^{(2)}$ as,
\eqabegin
Q^{(1)} &=& \left(e^{-\phi_0/2} \cos \theta + \chi_0 e^{\phi_0/2} \sin
\theta\right) \Delta_{(q_1, q_2)}^{1/2} Q\nn\\
Q^{(2)} &=& e^{\phi_0/2} \sin \theta \Delta_{(q_1, q_2)}^{1/2} Q
\eqaend
By demanding that the charges be quantized, we get from (30),
\eqabegin
\sin \theta &=& e^{-\phi_0/2} \Delta_{(q_1, q_2)}^{-1/2} q_2\nn\\
\cos \theta &=& e^{\phi_0/2} \left(q_1 - q_2 \chi_0\right) \Delta_
{(q_1, q_2)}^{-1/2}
\eqaend
where $q_1$ and $q_2$ are integers. Using $\cos^2 \theta + \sin^2 \theta
= 1$ we obtain from (31), the value of $\Delta_{(q_1, q_2)}$ to be
\eqabegin
\Delta_{(q_1, q_2)} &=& e^{-\phi_0} q_2^2 + \left(q_1 - q_2 \chi_0\right)
^2 e^{\phi_0}\nn\\
&=& \left(q_1,\,\, q_2\right) {\cal M}_{D0}^{-1} \left(\begin{array}{c}
q_1 \\ q_2\end{array}\right)
\eqaend
where ${\cal M}_{D0} = \left(\begin{array}{cc} \chi_0^2 + e^{-2\phi_0} &
\chi_0\\ \chi_0 & 1\end{array}\right)\,e^{\phi_0}$.
The important point to note here is that the expression for $\Delta_
{(q_1, q_2)}$ is SL(2, R) covariant\footnote[2]{As these string states
are BPS saturated, their tensions will not be renormalized and thus
the SL(2, Z) covariant formula of string tension gives a strong evidence
for the conjectured SL(2, Z) invariance in the theory.} and so is 
the charge as well as
the tension of a general string. This can be easily seen as the charge
$\left(\begin{array}{c} q_1\\q_2\end{array}\right)$ transforms as
$\Lambda \left(\begin{array}{c} q_1\\q_2\end{array}\right)$ under SL(2, R)
transformation. With the value of $\sin \theta$ and $\cos \theta$ in (31)
the transformed antisymmetric tensor field components can be obtained
from (21) as,
\eqabegin
B_{01}^{(1)} &=& e^{\phi_0}\left(q_1 - q_2 \chi_0\right) \Delta^{-1/2}_
{(q_1, q_2)} A^{-1}_{(q_1, q_2)}\nn\\
B_{01}^{(2)} &=& e^{\phi_0}\left(q_2 |\lambda_0|^2 - q_1 \chi_0\right)
\Delta_{(q_1, q_2)}^{-1/2} A^{-1}_{(q_1, q_2)}
\eqaend
which can be written compactly as follows,
\begineq
\left(\begin{array}{c} B_{01}^{(1)}\\B_{01}^{(2)}\end{array}\right)
= {\cal M}^{-1}_{D0} \left(\begin{array}{c}q_1\\q_2\end{array}\right)
\Delta^{-1/2}_{(q_1, q_2)} A^{-1}_{(q_1, q_2)}
\endeq
It can be checked that both sides of (34) has the right transformation
property under SL(2, R). Using (31), $\lambda$ can be calculated as 
follows:
\eqabegin
\lambda &=& \frac{q_1 \chi_0 - q_2 |\lambda_0|^2 + i q_1 e^{-\phi_0}
A^{1/2}_{(q_1, q_2)}}{q_1 - q_2 \chi_0 + i q_2 e^{-\phi_0} A^{1/2}_
{(q_1, q_2)}}\nn\\
&=& \frac{\chi_0 e^{-\phi_0} \Delta_{(q_1, q_2)} + q_1 q_2 e^{-2\phi_0}
\left(A_{(q_1, q_2)} - 1\right) + i \Delta_{(q_1, q_2)} A_{(q_1, q_2)}^
{1/2} e^{-2\phi_0}}{e^{-\phi_0} \Delta_{(q_1, q_2)} + q_2^2 e^{-2\phi_0}
\left(A_{(q_1, q_2)} -1 \right)}
\eqaend
So, from (35) we find the transformed value of $G_{mn}$ and $\Delta$ as
\eqabegin
G_{mn} &=& \left(\frac{e^{-\phi_0} \Delta_{(q_1, q_2)} + q_2^2 e^{-2\phi_0}
\left(A_{(q_1, q_2)} - 1\right)}{\Delta_{(q_1, q_2)} e^{-\phi_0}}\right)^
{4/(D-2)}\,\delta_{mn}\nn\\
\Delta &=& \left(\frac{e^{-\phi_0} \Delta_{(q_1, q_2)} + q_2^2 e^{-2\phi_0}
\left(A_{(q_1, q_2)} - 1\right)}{\Delta_{(q_1, q_2)} e^{-\phi_0}}\right)^
{2(10-D)/(D-2)}
\eqaend
Note that the generation of non-trivial $\Delta$ is consistent with the
SL(2, R) invariance of type II string effective action because the matrix
${\cal M}_D$ in Eq.(19) constructed to show the invariance in fact involves
$\tilde {\phi} = \phi + \frac{1}{2} \log \Delta$.

So, starting from the macroscopic string-like solution of Dabholkar
et. al. [13] in any $D <$ 10, we have been able to 
construct the SL(2, Z)
multiplets of string-like solutions of type II string theory given
by the metric in Eq.(4) and other field configurations in Eqs.(33)--
(36). This generalizes the construction of SL(2, Z) multiplets of
string solutions in $D = 10$ by Schwarz [4] to any $D <$ 10. 
Although formally
the expressions (33)--(35) look identical to the ten dimensional 
solutions of Schwarz, but they are completely different. This is because
the function $A_{(q_1, q_2)}$ is totally different for different $D$ 
as can be seen from Eq.(6). The solutions as we have seen are 
characterized by two integers $(q_1, q_2)$ corresponding to the charges
associated with $B_{01}^{(1)}$ and $B_{01}^{(2)}$.

Let us now discuss the stability of these solutions [6]. 
We have mentioned 
earlier that the charge and the tension of a general $(q_1, q_2)$
string is given by SL(2, Z) covariant expressions,
\eqabegin
\alpha_{(q_1, q_2)} &=& \Delta_{(q_1, q_2)}^{1/2} Q\,\,\,=\,\,\, \sqrt{
e^{-\phi_0} q_2^2 + \left(q_1 - q_2 \chi_0\right)^2 e^{\phi_0}} Q\\
T_{(q_1, q_2)} &=& \Delta_{(q_1, q_2)}^{1/2} T\,\,\,=\,\,\,\sqrt{e^{-\phi_0}
q_2^2 + \left(q_1 - q_2 \chi_0\right)^2 e^{\phi_0}} T
\eqaend
Using (37) and (38), it is easy to check that when $\chi = 0$, $(\alpha
_{(p_1, p_2)} + \alpha_{(q_1, q_2)})^2 \geq \alpha^2_{(p_1+q_1, 
p_2+q_2)}$ and similarly, $(T_{(p_1, p_2)} + T_{(q_1, q_2)})^2 \geq
T^2_{(p_1+q_1, p_2+q_2)}$. Since both $\alpha_{(q_1, q_2)}$ and
$T_{(q_1, q_2)}$ are positive real numbers, we conclude,
\eqabegin
\alpha_{(p_1, p_2)} + \alpha_{(q_1, q_2)} &\geq & 
\alpha_{(p_1+q_1, p_2+q_2)}\\
T_{(p_1, p_2)} + T_{(q_1, q_2)} &\geq & T_{(p_1+q_1, p_2+q_2)}
\eqaend
The equality holds when $p_1 q_2 = p_2 q_1$ or in other words when $p_1 =
n q_1$ and $p_2 = n q_2$ with $n$ being an integer. So, when $q_1, q_2$
are relatively prime, the inequality prevents the string from decaying
into multiple string states, as this configuration is 
energetically more
favorable than others. Eq.(40) is what we have called the tension gap
equation. Also, from (39) we note that when $q_1$,
$q_2$ are relatively prime, the charge conservation can not be satisfied
if the string breaks up into multiple strings. Thus the configuration
$(q_1, q_2)$, with $q_1$ and $q_2$ relatively prime, is perfectly stable
and denotes a bound state configuration [15] of $q_1$ fundamental 
strings with
$q_2$ D-strings [16].

To conclude, we have constructed in this paper the SL(2, Z) multiplets
of macroscopic string-like solutions of type II theory in any $D <$ 10.
This construction is made possible by a recent observation of the 
SL(2, R) invariance of toroidally compactified type IIB string effective
action. This generalizes the construction of SL(2, Z) multiplets of
string-like solutions of type IIB string theory in $D = 10$ by Schwarz.
Our solutions have formal similarity with the solutions in $D = 10$, 
but they are totally different as they involve dimensionally dependent
functions. The string-like solutions in $D <$ 10
 are also characterized by two relatively prime integers, as their 
counterpart in $D = 10$, corresponding to the charges of two antisymmetric
tensor fields in the theory. We have also discussed in brief the 
stability of the solutions from the charge conservation and tension 
gap relation. As we have mentioned earlier, there are more string-like
solutions not only with electric charge but also with magnetic charge 
in type II theories in lower dimensions which should form
multiplets of bigger symmetry group, the U-duality group. Apart from 
the string-like
solutions, there are also other $p$-brane solutions in these theories
which deserve a systematic study to properly identify the complete
U-duality group. This will provide strong evidence for the conjecture of
the U-duality symmetries in those theories. 

\vspace{1cm}

\begin{large}
\noindent{\bf Acknowledgements:}
\end{large}

\vspace{.5cm}

I am grateful to John Schwarz for very helpful correspondences. I would
also like to thank Jnanadeva Maharana for useful discussions.

\vspace{1cm}

\begin{large}
\noindent{\bf References:}
\end{large}

\vspace{.5cm}

\begin{enumerate}
\item B. Julia, in {\it Supergravity and Superspace}, Eds. S. W. Hawking
and M. Rocek, Cambridge University Press, 1981.
\item C. M. Hull and P. K. Townsend, \np 438 (1995) 109.
\item C. Hull, \pl 357 (1995) 545.
\item J. H. Schwarz, \pl 360 (1995) 13 (hep-th/9508143, last revision
in June, '97).
\item J. H. Schwarz, {\it Superstring Dualities}, hep-th/9509148.
\item J. H. Schwarz, {\it Lectures on Superstring and M-Theory
Dualities}, hep-th/9607201.
\item J. Maharana, {\it S-Duality and Compactification of Type IIB
Superstring Action}, hep-th/9703009.
\item S. Roy, {\it On S-Duality of Toroidally Compactified Type IIB
String Effective Action}, hep-th/9705016.
\item A. Giveon, M. Porrati and E. Rabinovici, Phys. Rep. C244 (1994)
77; E. Alvarez, L. Alvarez-Gaume and Y. Lozano, {\it An Introduction
to T-Duality in String Theory}, hep-th/9410237.
\item A. Sen, Int. J. Mod. Phys. A9 (1994) 3707.
\item M. J. Duff, R. R. Khuri and J. X. Lu, Phys. Rep. C259 (1995) 213.
\item A. Sen and C. Vafa, \np 455 (1995) 165.
\item A. Dabholkar, G. Gibbons, J. A. Harvey and F. Ruiz Ruiz, \np
340 (1990) 33; A. Dabholkar and J. A. Harvey, Phys. Rev. Lett. 63 (1989)
478.
\item N. Marcus and J. H. Schwarz, \pl 115 (1982) 111; J. H. Schwarz,
\np 226 (1983) 269; P. Howe and P. West, \np 238 (1984) 181.
\item E. Witten, \np 460 (1996) 335.
\item J. Polchinski, Phys. Rev. Lett. 75 (1995) 4724. 
\end{enumerate}
\vfil
\eject 

\end{document}